\documentclass[aps,pra,twocolumn,nofootinbib]{revtex4-1}
\usepackage{xcolor}
\usepackage{hyperref}
\usepackage{amssymb}
\usepackage{amsmath}
\usepackage{comment}
\usepackage{amsthm}
\newtheorem{lemma}{Lemma}
\usepackage{tikz-cd}
\newcommand{\eqnref}[1]{Eq.~(\ref{#1})}

\newcommand{\UU}{\mathrm{U}}
\newcommand{\ket}[1]{|#1\rangle}
\begin{document}
\title{Topological Goldstone phases of matter}
\date{\today}
\begin{abstract}
We consider the possibility for phases of matter in which a continuous symmetry is spontaneously broken in a \emph{topologically non-trivial} way, which, roughly, means that the action for the Goldstone modes contains a quantized topological term, and could manifest in, for example, non-trivial quantum numbers of topological defects of the order parameter. We show that, in fact, such a scenario can occur only when the system is in a non-trivial symmetry-protected topological (SPT) or symmetry-enriched topological (SET) phase with respect to the residual symmetry; or alternatively, if the original symmetry before spontaneous symmetry breaking acts on the system in an ``anomalous'' way. Our arguments are based on a general correspondence between topological defects of the order parameter and topological defects of a background gauge field for the residual symmetry.
\end{abstract}
\author{Dominic V. Else}
\affiliation{Department of Physics, Harvard University, Cambridge, MA 02138, USA}
\affiliation{Department of Physics, Massachusetts Institute of Technology, Cambridge, MA 02139, USA}
\maketitle

Zero-temperature phases of matter are traditionally divided into two categories: spontaneous symmetry breaking phases, and topological phases which are characterized in terms of ``quantum topology'' of their ground state, that is not related to spontaneous symmetry breaking of any symmetry \cite{Wen_1610}. By contrast, in this work, we will discuss phases of matter in which there is a non-trivial interplay between spontaneous symmetry breaking and quantum topology.

We consider a system which originally has symmetry $G$, which is broken down to a subgroup $H$ by spontaneous symmetry breaking. Then there is an order parameter that lives in the quotient space $\Lambda := G/H$, and we have a manifold of degenerate symmetry-breaking ground states labelled by the order parameter. Moreover, if the order parameter space $\Lambda$ is not discrete, then the ground state will be gapless (in addition to the aforementioned degeneracy) due to the Goldstone modes. In this paper, we consider scenarios where the action that describes these Goldstone modes can contain a \emph{quantized topological term}, reflecting a non-trivial quantum topology associated with these modes.

Our main goal in this paper will be to give the general theory of when the Goldstone modes of a spontaneous symmetry breaking phase carry non-trivial quantum topology. Our main result is that, provided the original symmetry $G$ is not anomalous [that is, provided that the symmetry is microscopically represented in an on-site manner without requiring the system to be at the boundary of a higher-dimensional symmetry-protected topological (SPT) phase], the quantum topology of the Goldstone modes can always be related back to the SPT or symmetry-enriched topological (SET) phase carried by the spontaneous symmetry-breaking ground state with respect to the residual symmetry $H$. (For technical reasons, we will give the full derivation of this result only for bosonic systems with only unitary symmetries, but we expect it to hold with anti-unitary symmetries and in fermionic systems as well.) Some special cases of this general observation have previously been studied in Ref.~\cite{Brauner_1405}.

The outline of the remainder of this paper is as follows. In Section \ref{sec:first_example} we will discuss a simple example of a spontaneous symmetry-breaking phase of matter in which the Goldstone modes have non-trivial quantum topology. In Section \ref{sec:definition}, we will give the precise and general definition of quantum topology of Goldstone modes. In Section \ref{sec:relation}, we discuss a relation between topological defects of the Goldstone modes and the topological defects of $H$ gauge fields, which is already suggestive that the quantum topology of the Goldstone modes is determined by the SPT/SET order with respect to the residual symmetry $H$. In Section \ref{sec:examples} we apply this correspondence to some examples. In Section \ref{sec:anomalous} we will discuss how the situation could change if the symmetry is anomalous. Finally, in Section \ref{sec:equivalence} (supplemented by Appendix \ref{appendix:formal_theory}) we give the formal theory in complete generality in order to justify all the claims made in the paper, and then conclude in Section \ref{sec:conclusion}.

\section{A simple example: quantum Hall ferromagnet}
\label{sec:first_example}
A quantum Hall ferromagnet \cite{Sondhi_1993} is a system of electrons in $d=2$ dimensions which initially has $G = \mathrm{U}(2)$ symmetry [i.e. it contains spin rotation $\mathrm{SU}(2)$ and charge conservation $\UU(1)$ as subgroups], which is spontaneously broken down to $\UU(1)_{\uparrow} \times \UU(1)_\downarrow$ by anti-ferromagnetic order, where the subgroup $\UU(1)_\uparrow \times \UU(1)_\downarrow$ corresponds to the diagonal unitary matrices if we view $\UU(2)$ as a matrix group (which is to say that the charges of $\UU(1)_\uparrow$ and $\UU(1)_\downarrow$ correspond to the number of up-spin and down-spin electrons respectively.) The  order parameter lives in the manifold $\Lambda = G/H \cong S^2$, that is it can be represented as a three-dimensional unit vector $\mathbf{n}$.

The quantum Hall ferromagnet has the property that a skyrmion of the order parameter carries unit electric charge. We consider this to be a manifestation of the non-trivial quantum topology of the Goldstone modes. Another way to say this is that if we write the action describing the long-wavelength dynamics of the Goldstone modes coupled to a background electromagnetic gauge field $A_\mu$ [i.e. a gauge field for the charge $\UU(1)$, which is the diagonal subgroup of the residual group $H = \mathrm{U}(1)_\uparrow \times \mathrm{U}(1)_\downarrow$], it will contain a topological term, namely:
\begin{equation}
\label{eq:qsh_topological_term}
S = S_0 + \frac{1}{4\pi} \int d^3 x \epsilon_{\mu \nu \lambda} A_\mu \epsilon^{a b c} n^a  \partial_\nu n^b \partial_\mu n^c,
\end{equation}
where $S_0$ is the action in the absence of applied electromagnetic field.
(One can verify that the topological term is invariant, modulo $2\pi$, under gauge transformations, including large gauge transformations, due to the quantization of the skyrmion number.) A consequence of these properties is, for example, that if we restore spin rotation symmetry by condensing pairs of skyrmions (since the elementary skyrmions carry charge $1$, they are fermionic and must be condensed in pairs), then the charge $\mathrm{U}(1)$ symmetry will be spontaneously broken.

\section{General definition of quantum topology of Goldstone modes}
\label{sec:definition}
Let us now give a precise definition of quantum topology of Goldstone modes which formalizes the idea of the action of the Goldstone modes containing a topological term. The Goldstone modes by definition are weakly fluctuating and essentially classical since otherwise the symmetry would be restored. Therefore, in order to characterize the quantum topology, it should be legitimate to add a small symmetry-breaking field to gap out the Goldstone modes, leaving behind a gapped quantum ground state where any remaining degeneracy is topological and corresponds to ground states that are locally indistinguishable. Since the effect of this field is essentially to pin the order parameter to a specific value, we can imagine that the symmetry-breaking field is itself parameterized by the order parameter space $\Lambda$. Therefore, what we eventually obtain is a \emph{family} of gapped ground states parameterized by $\Lambda$.
Furthermore, let $H_0$ be the largest subgroup of $G$ that acts trivially on the whole order parameter space $\Lambda$ (it can equivalently be defined as the subgroup of $H$ comprising group elements $h$ such that $g h g^{-1} \in H$ for all $g \in G$.) Then each ground state in the family is invariant under $H_0$, so we can also talk about $H_0$-enriched families of gapped ground states. If $H$ is a normal subgroup of $G$, then $H_0$ is just equal to $H$. By contrast, for the quantum Hall ferromagnet example above, then then $H_0$ is only the diagonal subgroup of $H = \mathrm{U}(1)_\uparrow \times \mathrm{U}(1)_\downarrow$, corresponding to the total electron number.

It is by now well understood that such families can exhibit non-trivial topological winding in the space of all gapped $H_0$-symmetric states \cite{Thorngren_1612, Thorngren_1710, Cordova_1905_a, Cordova_1905_b, Kapustin_2001, Kapustin_2003, Hsin_2004}. A manifestation of such non-trivial winding will be topological terms [such as \eqnref{eq:qsh_topological_term}] in the action for Goldstone modes coupled to a background gauge field of $H_0$.
The non-trivial $H_0$-enriched winding can also have consequences for the properties of topological excitations and defects, as we already saw in the example above. In the SPT case, in general a smooth topological excitation such as the skyrmion considered above can carry non-trivial $H_0$ charge, while a topological defect where the order parameter field becomes singular on a $k$-dimension submanifold can carry protected gapless modes analogous to those that occur at the boundary of an $H_0$ SPT phase in $k+1$ spatial dimensions.


\section{Relation to the SPT/SET order of the symmetry-breaking ground state}
\label{sec:relation}
In the example discussed above, it is the case that if we fix the order parameter to a specific value and look at the symmetry-breaking ground state, then it is in a non-trivial SPT phase (specifically, a quantum spin Hall phase) with respect to the residual symmetry $H$. One might suspect that the topological term carried by the Goldstone modes is in fact a consequence of this. We will show that, indeed, provided that the original symmetry $G$ is not anomalous, the topological non-triviality of the Goldstone modes must come from, and can be computed by means of, the SPT or SET order of the symmetry-breaking ground state with respect to the residual symmetry group $H$.

We will give the general theory behind this statement in later sections. Here we just want to emphasize a quick way to understand this result, at least in some special cases.
This comes by considering the long exact sequence of homotopy groups associated with the fiber bundle $G \to \Lambda$ \cite{Hatcher}, namely:
\begin{multline}
\label{eq:long_exact_sequence}
\cdots \to \pi_2(H) \to \pi_2(G) \to \pi_2(\Lambda) \\
\to \pi_1(H) \to \pi_1(G) \to \pi_1(\Lambda) \\
\to \pi_0(H) \to \pi_0(G) \to \pi_0(\Lambda)
\end{multline}
Of particular interest are the so-called ``connecting homomorphisms'' $\pi_k(\Lambda) \to \pi_{k-1}(H)$. We know \cite{Mermin_1979} that topological defects of the order parameter where the order parameter becomes singular on a manifold of codimension $k$ are classified by $\pi_k(\Lambda)$; for example, in three spatial dimensions, if $G = \mathrm{SU}(2), H = \mathrm{U}(1)$, then $\Lambda = S^2$ and $\pi_2(\Lambda) = \mathbb{Z}$ classifies the so-called ``hedgehog defects'' in three spatial dimensions. Meanwhile, we also know \cite{Atiyah_1978} that gauge field defects that are sourced by objects of codimension $k$ are classified by $\pi_{k-1}(H)$; for example if $H = \mathrm{U}(1)$, then $\pi_1(\mathrm{U}(1)) = \mathbb{Z}$ classifies monopoles in three spatial dimensions. Therefore, what we have found is that there is a mapping from order parameter defects into gauge field defects for the residual symmetry; for example, the hedgehog defects map into monopoles in the example just described.

This mapping is perhaps most readily understood from a physical point of view if we imagine gauging the $G$ symmetry and coupling to a dynamical $G$ gauge field. Then the order parameter acts as a ``Higgs field'' such that the ``effective gauge group'' at low energies is reduced to $H$. In this gauge theory, the spatial variation of the order parameter can be gauged away, but the topological defects still survive as defects of the $H$ gauge field. 

What we will eventually argue is that a similar correspondence still survives even without gauging the $G$ symmetry, at the level of the topological properties of the defects. In particular, let $H_{\bullet}$ be the intersection of $H_0$ with the center of $H$. (For the concrete examples we discuss in this paper, $H$ is always Abelian so $H_\bullet = H_0$). The point of defining $H_\bullet$ is that it is the largest subgroup of $H_0$ that does not risk getting broken by an $H$ gauge field configuration when $H$ is non-Abelian.

We will derive the following physical statement: the $H_\bullet$-enriched properties of a topological defect of the order parameter (for example, the $H_\bullet$ charge of a skyrmion or anomalous $H_\bullet$ action on the core of a vortex) will be the \emph{same} as the $H_\bullet$-enriched properties of the symmetry-breaking ground state coupled to the corresponding topological defect of the $H$ gauge field. The latter are always determined by the SPT/SET order with respect to $H$ of the symmetry-breaking ground state. 


\section{Application of the general formalism to some examples}
\label{sec:examples}
\subsection{Quantum Hall ferromagnet}
Consider the quantum Hall ferromagnet example that we introduced in Section \ref{sec:first_example}. The residual symmetry is $H = \mathrm{U}(1)_\uparrow \times \mathrm{U}(1)_\downarrow$, while $H_0$ is the diagonal subroup (generated by the total electron number). The first step is identify the SPT phase carried with respect to the residual symmetry $H$.  For the quantum Hall ferromagnet, this phase is the one which contains a state where fully spin-polarized electrons (i.e.\ all spin-up) form a $\nu=1$ quantum Hall state.

An elementary skyrmion corresponds to the generator of the homotopy group $\pi_2(S^2) = \mathbb{Z}$. We want to determine what gauge field configuration of $H = \mathrm{U}(1)_\uparrow \times \mathrm{U}(1)_\downarrow$ this maps into. We can invoke the exact sequence \eqnref{eq:long_exact_sequence}. We know that $\pi_2(\mathrm{U}(2)) = 0$ and $\pi_1(S^2) = 0$, so we obtain the short exact sequence
\begin{equation}
\begin{tikzcd}[column sep=0.4cm]
0 \arrow[r] & \pi_2(S^2) \arrow[r] \arrow[d,equal] & \pi_1(\mathrm{U}(1) \times \mathrm{U}(1)) \arrow[r] \arrow[d,equal] & \pi_1(\mathrm{U}(2)) \arrow[r] \arrow[d,equal] & 0 \\
 & \mathbb{Z} & \mathbb{Z} \times \mathbb{Z} & \mathbb{Z} \\
\end{tikzcd}
\end{equation}
From this exact sequence it is clear that the elementary skyrmion maps into the combination of a $2\pi$ flux of $\mathrm{U}(1)_\uparrow$ and a $-2\pi$ flux of $\mathrm{U}(1)_\downarrow$. This object carries unit electric charge (as one can easily see by considering the fully-spin-polarized quantum Hall state). Therefore, we find that skyrmions carry unit charge, as expected.

\subsection{Fermionic skyrmions in superfluids}
The second example we consider is closely related to the quantum Hall ferromagnet above, but we introduce superfluid order (corresponding to condensing electron pairs), so that the residual group is only a single $\mathrm{U}(1)_{z}$ generated by the $z$-component of the spin.
Thus, we have $G = \mathrm{U}(2)$ and $H = \mathrm{U}(1)_{z}$, while $H_0$ is the $Z_2$ subgroup of $\mathrm{U}(1)_z$, whose generator can be interpreted as the fermion parity. Therefore we will consider under which circumstances the skyrmions become fermionic.

We obtain the exact sequence
\begin{equation}
\begin{tikzcd}[column sep=0.4cm]
0 \arrow[r] & \pi_2(\Lambda) \arrow[r] & \pi_1(\mathrm{U}(1)_{z}) \arrow[r] \arrow[d,equal] & \pi_1(\mathrm{U}(2)) \arrow[d,equal] \\
 &  & \mathbb{Z} & \mathbb{Z} \\
\end{tikzcd}
\end{equation}
The mapping of $\mathrm{U}(1)_z$ into $\mathrm{U}(2)$ is such that every element of $\pi_1(\mathrm{U}(1)_z)$ maps to zero in the corresponding map $\pi_1(\mathrm{U}(1)_z) \to \pi_1(\mathrm{U}(2))$. Hence from the above exact sequence we can conclude that $\pi_2(\Lambda) \cong \mathbb{Z}$ (corresponding to the skyrmions of the ferromagnetism), and an elementary skyrmion maps into an elementary flux quantum for $\mathrm{U}(1)_z$.

Now assume that the system has a spin quantum Hall conductance $\sigma$ (in units where spin quantum Hall conductance is quantized to integers). Then an elementary flux quantum for $\mathrm{U}(1)_z$ carries spin $S^z = \sigma/2$, and in particular it has fermion parity $(-1)^{\sigma}$. Thus, we conclude that if $\sigma$ is odd, then a skyrmion is fermionic. In fact, this can be described by a Hopf term in the action for the order parameter \cite{Wilczek_1983}. This behavior was proposed in Ref.~\cite{Volovik_1989} to occur in ${}^3 \mathrm{He}$ films, with electrons replaced by the fermionic ${}^3 \mathrm{He}$ atoms.

\subsection{Fractional fermion parity of hedgehogs in a spinful topological superconductor}
Our third example is a superconductor in $d=3$ spatial dimensions which originally preserves the $\mathrm{SU}(2)$ spin rotation symmetry as well as time-reversal symmetry, which squares to the fermion parity $(-1)^F$ [which is the non-trivial central element of $\mathrm{SU}(2)$] as is appropriate for an electronic system. These generate the original symmetry group $G$. Then the symmetry gets spontaneously broken down to the group $H$ generated by a $\mathrm{U}(1)$ subgroup of $\mathrm{SU}(2)$ and by time-reversal (while $H_0 = Z_4^t$, generated by time-reversal.)
  Thus, we are considering an exotic kind of ``ferromagnetic'' order where the order parameter is a vector, but it is not identified with magnetization, but rather is even under time reversal symmetry, so that time reversal symmetry is preserved. 
  
The interacting topological phases with respect to the residual symmetry group $H$ in this example have been classified in Ref.~\cite{Wang_1401}, resulting in a $\mathbb{Z}_8$ classification (see the ``class AIII'' line of Table I of Ref.~\cite{Wang_1401}). Let us suppose that we set up the system so that the symmetry-breaking ground states are in the root phase of this classification with respect to $H$. This phase exhibits the Witten effect, which is to say  that a monopole of the charge $\mathrm{U}(1)$ symmetry [the diagonal subgroup of $\UU(1)_\uparrow \times \UU(1)_\downarrow$] carries \emph{fractional} charge.

The order parameter lives in $\Lambda = S^2$, so we can consider a hedgehog defect corresponding to the generator of $\pi_2(S^2) = \mathbb{Z}$. Then, since we have that $\pi_2(G) = \pi_1(G) = 0$,  we obtain the exact sequence
\begin{equation}
\begin{tikzcd}[column sep=0.4cm]
0 \arrow[r] & \pi_2(S^2) \arrow[r] \arrow[d,equal] & \pi_1(H) \arrow[r] \arrow[d,equal] & 0 \\
 & \mathbb{Z} & \mathbb{Z} \\
\end{tikzcd}
\end{equation}
which shows that an elementary  hedgehog defect maps into an elementary monopole of the residual $\mathrm{U}(1)$ symmetry. Due to the Witten effect, this defect therefore acquires a fractional charge. Note that because the full $\mathrm{U}(1)$ is not a subgroup of $H_0$, the fractional charge of the hedgehog defect under the full $\mathrm{U}(1)$ is not well-defined, but $H_0$ does contain the fermion parity $(-1)^F$, so it \emph{is} still well-defined to say that the hedgehog defect carries fractional charge under the fermion parity. This corresponds to a non-trivial projective representation of the $Z_4^t$ group generated by time-reversal symmetry, hence there is at least a two-fold degeneracy associated with the hedgehog defect.

\section{Anomalous symmetries}
\label{sec:anomalous}
Above, we made the assumption that the low-energy physics described occurs in a lattice system where $G$ acts on an on-site manner so that there is no anomaly. In such case, the result of this paper is that non-trivial quantum topology of the Goldstone modes can only originate from the SPT/SET order of the symmetry-breaking ground state with respect to the residual symmetry $H$. In particular, it follows that if $G$ is spontaneously broken all the way to nothing, so $H = 1$, then there is no non-trivial quantum topology of the Goldstone modes.

If $G$ is spontaneously broken all the way to nothing, then the order parameter manifold $\Lambda$ is $G$ itself.
Recall that if the manifold dimension of $G$ is $d+2$, where $d$ is the spatial dimension, then then there is a non-trivial $\mathbb{Z}$-valued invariant for families of ground states parameterized by $G$ \cite{Kapustin_2003}. For example, if we set $d=1$, $G=\mathrm{SU}(2)$, $H=1$, [and disregard the fact that dynamical fluctuations destroy spontaneous symmetry breaking in this dimension, which is not relevant for the present discussion; one can instead just treat the order parameter field as a classical background], one might have expected that the action could contain a Wess-Zumino-Witten term \cite{Witten_1984}:
\begin{equation}
S = S_0 + \frac{1}{24\pi} \int_B \mathrm{Tr} [g^{-1} dg \wedge g^{-1} dg \wedge g^{-1} dg],
\end{equation}
where $g \in \mathrm{SU}(2)$ is the order parameter field, which has been extended to a three-dimensional manifold $B$ for which the physical two-dimension space-time is the boundary.
However, what we have found is that such a term, or its higher-dimensional generalizations, \emph{cannot} appear in this scenario if the $G$ symmetry is non-anomalous.

On the other hand, it is well known that such a Wess-Zumino-Witten term does appear on the \emph{boundary} of a bosonic $\mathrm{SU}(2)$ SPT in $d=2$ spatial dimensions \cite{Liu_1205}, in which case the symmetry action on the 1-dimensional boundary is ``anomalous'' and cannot be represented in an on-site manner. This demonstrates that when the symmetry that is spontaneously broken is anomalous, then it is possible to have non-trivial quantum topology of the Goldstone modes that is not linked to the SPT/SET order of the symmetry-breaking ground state with respect to the residual symmetry.

\section{The general theory of equivalence}
\label{sec:equivalence}
In this section, we will outline the general reason why the SPT/SET order of the residual symmetry fully determines the quantum topology of the Goldstone modes. The basic idea is the following. For an SPT or SET phase with symmetry $G$, one probes the SPT/SET order by considering the response of the system to a topological class of $G$ gauge fields for the $G$ symmetry, described by a principal $G$-bundle on $M$. 
(For fermionic systems and bosonic systems with anti-unitary symmetries, the precise definition of a ``$G$ gauge field'' is a bit more complicated. We will confine our derivations in what follows to bosonic systems with only unitary symmetries, but we expect that similar results will hold more generally.)

We will refer to the set of principal $G$-bundles on $M$ as $\mathrm{Gauge}_G(M)$. Meanwhile, the generalization of this idea for $G$-enriched families of states parameterized by a space $\Lambda$ is to consider the response of the system to some topological class of parameter configurations on the space-time $M$, described by a homotopy class of maps $\lambda : M \to \Lambda$, along with a principal $G$-bundle on $M$. We will call the set of such data $\mathrm{Gauge}_{G;\Lambda}(M)$.

What we will show is that in the case of spontaneous symmetry breaking, where $\Lambda = G/H$, the response of the system to $\mathrm{Gauge}_{H_0;\Lambda}(M)$ is completely determined by the response to $\mathrm{Gauge}_{H}(M)$, where as before $H_0$ is the largest normal subgroup of $G$ that is a subgroup of $H$. The point is that, mathematically, there is a natural map
\begin{equation}
\label{eq:gauge_mapping}
f : \mathrm{Gauge}_{H_0;\Lambda}(M) \to \mathrm{Gauge}_{H}(M),
\end{equation}
We show in Appendix \ref{appendix:formal_theory} that the physical response of the system to an element $\mathcal{G} \in \mathrm{Gauge}_{H_0;\lambda}(M)$ is indeed equivalent to the response to $f(\mathcal{G})$. Here we will content ourselves with constructing the map $f$.

A rough physical interpretation of the map can be expressed in terms of the picture mentioned in Section \ref{sec:relation} where we gauge the $G$ symmetry and couple to a dynamical $G$ gauge field. In this case, the map \eqnref{eq:gauge_mapping} will simply describe a gauge transformation into the so-called ``unitary gauge'' where the spatial variation of the order parameter is eliminated and the remaining part of the gauge field that does not acquire a mass through the Higgs mechanism constitutes an $H$ gauge field.

To construct the map, we first need to recall the definition of a principal $G$-bundle on a space $M$. We consider an open covering of $M$; that is, a collection of open patches $U_i \subseteq M$ such that $\cup_i U_i = M$. A principal $G$-bundle on $M$ is specified by the transition functions, which are continuous maps $g_{ij} : U_i \cap U_j \to G$, satisfying the following conditions:
\begin{align}
g_{ii}(x) &= 1 \quad (x \in U_i) \\
g_{ij}(x) &= g_{ji}(x)^{-1} \quad (x \in U_i \cap U_j), \\
g_{ik}(x) &= g_{ij}(x) g_{jk}(x) \quad (x \in U_i \cap U_j \cap U_k),
\end{align}
We identify principal bundles that are related by continuous functions $\phi_i(x) : U_i \to G$ such that $g_{ij}'(x) = \phi_i(x) g_{ij}(x) \phi_j(x)^{-1}$; by
replacing two patches $U_i$ and $U_j$, such that $g_{ij}(x) = 1$ for all $x \in U_i \cap U_j$, with their union $U_i \cup U_j$; or by homotopies (i.e. continuous deformations) of the transition functions.

The space $\Lambda$ is acted upon by the group $G$. By definition, for any point $\lambda \in \Lambda$, the subgroup of $G$ that leaves $\lambda$ fixed is in the same conjugacy class as $H$. Let us arbitrarily choose some $\lambda_* \in \Lambda$, and we adopt the convention that the subgroup that leaves $\lambda_*$ fixed is $H$ itself (not just a member of the same conjugacy class).

Now consider an element of $\mathrm{Gauge}_{H_0;\Lambda}(M)$. It is specified by a map $\lambda : M \to \Lambda$ and a principal $H_0$-bundle on $M$ specified by patches $\{ U_i \}$ and transition functions $h^{(0)}_{ij} : U_i \cap U_j \to H_0$. Now on each patch $U_i$, we look for a function $\varphi_i : U_i \to G$ such that $\lambda(x) = \varphi_i(x) \lambda_*$ for $x \in U_i$.
 (This may require splitting the patches up into smaller ones). Then we define the transition functions for a principal $H$-bundle on $M$ according to
\begin{equation}
h_{ij}(x) = \varphi_i(x)^{-1} h^{(0)}_{ij}(x) \varphi_j(x)
\end{equation}
Note that $h_{ij}(x) \in H$ because
\begin{align}
h_{ij}(x) \lambda_* &= \varphi_i(x)^{-1} h^{(0)}_{ij} \varphi_j(x) \lambda_* \\
&= \varphi_i(x)^{-1} h^{(0)}_{ij} \lambda(x) \\
&= \varphi_i(x)^{-1} \lambda(x) \\
&= \lambda_*.
\end{align}
Moreover, although there is some ambiguity in defining the functions $\varphi_i$, this does not affect the definition of the principal $H$-bundle up to gauge equivalence because two such choices $\varphi_i$ and $\varphi_i'$ are related by $\varphi_i'(x) = \varphi_i(x) \phi_i(x)$ for some function $\phi_i : U_i \to H$, such that the new transition functions are given by $h_{ij}'(x) = \phi_i(x)^{-1} h_{ij}(x) \phi_j(x)$.

Thus, we have finished giving a construction of the map \eqnref{eq:gauge_mapping}.

\section{Conclusion}
\label{sec:conclusion}
In this work we have given the general theory of quantum topology of the Goldstone modes in the case where the symmetry does not have an anomaly. It remains an interesting question to consider in more generality what happens when the symmetry \emph{does} have an anomaly, i.e. the system exists at the boundary of an SPT phase in one higher dimension.

Although in this work we have focussed on internal symmetries, similar considerations apply for spatial symmetries, such as translational symmetry, in light of the ``crystalline equivalence principle'' of Ref.~\cite{Thorngren_1612}. Some implications for crystalline and quasicrystalline systems will be explored in a subsequent work \cite{ToAppear}.

\begin{acknowledgments}
I thank Ying Ran, Ryan Thorngren, and T.~Senthil for helpful discussions. I thank Grigory Volovik for pointing out a reference. I thank the anonymous referee for suggesting that the map of Section \ref{sec:equivalence} could be viewed as a gauge transformation in a $G$ gauge theory. The author was supported by the EPiQS
Initiative of the Gordon and Betty Moore Foundation,
Grant Nos.~GBMF8683 and GBMF8684.
\end{acknowledgments}

\appendix
\section{The formal theory}
\label{appendix:formal_theory}
Here we will give the formal argument for the proposition that the quantum topology of the Goldstone modes descends from the SPT/SET order of the residual symmetry.
We employ the homotopy-theoretic point of view originally due to Kitaev \cite{Kitaev_0506, KitaevIPAM}, who showed \cite{Kitaev_0506} that any gapped ground state in $d$ spatial dimensions that is invariant under an on-site representation of some symmetry group $G$ defines a continuous map from $BG \to \Omega_d$ (we are again assuming a bosonic system with only unitary symmetries.) Here $BG = EG/G$ is the so-called ``classifying space'' of $G$, obtained by taking the quotient by the $G$ action of some contractible space $EG$ with a free action of $G$; and $\Omega_d$ is the space of all gapped ground states in $d$ spatial dimensions. One then conjectures that SPT and SET phases are precisely classified by homotopy classes of maps $BG \to \Omega_d$, and indeed all known classification schemes can be understood in terms of this conjecture by making specific assumptions about the topology of the space $\Omega_d$.

We can generalize the perspective of Kitaev to classify \emph{families} of states \cite{Thorngren_1612}. Let $\Lambda$ be some space with a $G$-action, and consider a family $\ket{\Psi_\lambda}, \lambda \in \Lambda$ of gapped ground states in $d$ spatial dimensions, such that $U(g) \ket{\Psi_\lambda} = \ket{\Psi_{g\lambda}}$, where $U(g)$ is an on-site representation of $G$. Then one can generalize the argument of Ref.~\cite{Kitaev_0506} to show that this defines a map from $\Lambda // G \to \Omega_d$, where we have defined
\begin{equation}
\Lambda // G := (\Lambda \times EG)/G,
\end{equation}
where $EG$ is as before, and the $G$ action on $\Lambda \times EG$ is defined in terms of the $G$ actions on $\Lambda$ and on $EG$. Therefore, we expect that the classification of such families is precisely given by homotopy classes of maps
\begin{equation}
\label{eq:Psi_map}
\Psi : \Lambda // G \to \Omega_d.
\end{equation}
In particular since there is a natural projection map\footnote{To see this, note that since $EG$ is itself a contractible space on which $H_0$ acts freely, we can take  $EH_0 = EG$. Then we have that $\Lambda \times BH_0 = \Lambda \times (EG/H_0) = (\Lambda \times EG)/H_0$ (since $H_0$ acts trivially on $\Lambda$), and this comes with a natural projection map into $\Lambda // G = (\Lambda \times EG)/G$.} $\Lambda \times BH_0 \to \Lambda // G$ (where as earlier, $H_0$ is the subgroup of $H$ that acts trivially on the whole space $\Lambda$), a map of the form \eqnref{eq:Psi_map} contains all the data of $H_0$-enriched topological families parameterized by $\Lambda$, such as, for example, the $H_0$ charge of topological defects.

Let us now argue that the homotopy class of the map $\Psi$ of \eqnref{eq:Psi_map} is already fully determined by the SPT/SET order with respect to $H$ carried by a single symmetry-breaking ground state. Indeed, this follows from the following lemma:
\begin{lemma}
Let $\Lambda$ be a space with a transitive $G$-action. Fix some $\lambda_* \in \Lambda$, and let $H$ be the subgroup of $G$ that leaves $\lambda_*$ fixed. (Thus, we can identify $\Lambda = G/H$.)
Then there is a homotopy equivalence between $\Lambda // G$ and $BH$.
\begin{proof}
Up to homotopy equivalence we can write $BH = EG/H$, since $EG$ is itself a contractible space on which $H$ acts freely. Then we will show that there is a homeomorphism between $\Lambda // G$ and $BH$. We need to construct an invertible continuous map from $\Lambda // G$ to $BH$ and show that its inverse is also continuous.

We define the map
\begin{equation}
\xi : \Lambda // G \to BH, [(\lambda, e)] \mapsto \pi(g_\lambda^{-1} e),
\end{equation}
where $(\lambda,e) \in \Lambda \times EG$ is a representative of the class $[(\lambda,e)] \in (\Lambda \times EG)/G$;
$\pi : EG \to BG$ is the projection map; and $g_\lambda$ is some element of $G$ such that $\lambda = g_\lambda \lambda_*$. One can show that $\xi$ is continuous and does not depend on the arbitrary choice of $g_\lambda$, nor on the choice of representative $(\lambda,e)$.

Meanwhile, we define the map
\begin{equation}
\chi : BH \to \Lambda // G, [e] \mapsto \pi'((\lambda_*,e))
\end{equation}
where $e \in EG$ is a representative for the class $[e] \in EG/H = BH$, and $\pi' : \Lambda \times EG \to \Lambda // G$ is the projection map. One can show that $\chi$ is continuous and independent of the choice of representative $e$.

One readily verifies that $\xi$ and $\chi$ are inverses of each other, and the result follows.

\end{proof}
\end{lemma}

\bibliography{ref-autobib,ref-manual}

\begin{thebibliography}{21}%
\makeatletter
\providecommand \@ifxundefined [1]{%
 \@ifx{#1\undefined}
}%
\providecommand \@ifnum [1]{%
 \ifnum #1\expandafter \@firstoftwo
 \else \expandafter \@secondoftwo
 \fi
}%
\providecommand \@ifx [1]{%
 \ifx #1\expandafter \@firstoftwo
 \else \expandafter \@secondoftwo
 \fi
}%
\providecommand \natexlab [1]{#1}%
\providecommand \enquote  [1]{``#1''}%
\providecommand \bibnamefont  [1]{#1}%
\providecommand \bibfnamefont [1]{#1}%
\providecommand \citenamefont [1]{#1}%
\providecommand \href@noop [0]{\@secondoftwo}%
\providecommand \href [0]{\begingroup \@sanitize@url \@href}%
\providecommand \@href[1]{\@@startlink{#1}\@@href}%
\providecommand \@@href[1]{\endgroup#1\@@endlink}%
\providecommand \@sanitize@url [0]{\catcode `\\12\catcode `\$12\catcode
  `\&12\catcode `\#12\catcode `\^12\catcode `\_12\catcode `\%12\relax}%
\providecommand \@@startlink[1]{}%
\providecommand \@@endlink[0]{}%
\providecommand \url  [0]{\begingroup\@sanitize@url \@url }%
\providecommand \@url [1]{\endgroup\@href {#1}{\urlprefix }}%
\providecommand \urlprefix  [0]{URL }%
\providecommand \Eprint [0]{\href }%
\providecommand \doibase [0]{http://dx.doi.org/}%
\providecommand \selectlanguage [0]{\@gobble}%
\providecommand \bibinfo  [0]{\@secondoftwo}%
\providecommand \bibfield  [0]{\@secondoftwo}%
\providecommand \translation [1]{[#1]}%
\providecommand \BibitemOpen [0]{}%
\providecommand \bibitemStop [0]{}%
\providecommand \bibitemNoStop [0]{.\EOS\space}%
\providecommand \EOS [0]{\spacefactor3000\relax}%
\providecommand \BibitemShut  [1]{\csname bibitem#1\endcsname}%
\let\auto@bib@innerbib\@empty
\bibitem [{\citenamefont {Wen}(2017)}]{Wen_1610}%
  \BibitemOpen
  \bibfield  {author} {\bibinfo {author} {\bibfnamefont {X.-G.}\ \bibnamefont
  {Wen}},\ }\href {\doibase 10.1103/RevModPhys.89.041004} {\bibfield  {journal}
  {\bibinfo  {journal} {Rev. Mod. Phys.}\ }\textbf {\bibinfo {volume} {89}},\
  \bibinfo {pages} {041004} (\bibinfo {year} {2017})},\ \Eprint
  {http://arxiv.org/abs/1610.03911} {arXiv:1610.03911} \BibitemShut {NoStop}%
\bibitem [{\citenamefont {Brauner}\ and\ \citenamefont
  {Moroz}(2014)}]{Brauner_1405}%
  \BibitemOpen
  \bibfield  {author} {\bibinfo {author} {\bibfnamefont {T.}~\bibnamefont
  {Brauner}}\ and\ \bibinfo {author} {\bibfnamefont {S.}~\bibnamefont
  {Moroz}},\ }\href {\doibase 10.1103/PhysRevD.90.121701} {\bibfield  {journal}
  {\bibinfo  {journal} {Phys. Rev. D}\ }\textbf {\bibinfo {volume} {90}},\
  \bibinfo {pages} {121701} (\bibinfo {year} {2014})},\ \Eprint
  {http://arxiv.org/abs/1405.2670} {arXiv:1405.2670} \BibitemShut {NoStop}%
\bibitem [{\citenamefont {Sondhi}\ \emph {et~al.}(1993)\citenamefont {Sondhi},
  \citenamefont {Karlhede}, \citenamefont {Kivelson},\ and\ \citenamefont
  {Rezayi}}]{Sondhi_1993}%
  \BibitemOpen
  \bibfield  {author} {\bibinfo {author} {\bibfnamefont {S.~L.}\ \bibnamefont
  {Sondhi}}, \bibinfo {author} {\bibfnamefont {A.}~\bibnamefont {Karlhede}},
  \bibinfo {author} {\bibfnamefont {S.~A.}\ \bibnamefont {Kivelson}}, \ and\
  \bibinfo {author} {\bibfnamefont {E.~H.}\ \bibnamefont {Rezayi}},\ }\href
  {\doibase 10.1103/PhysRevB.47.16419} {\bibfield  {journal} {\bibinfo
  {journal} {Phys. Rev. B}\ }\textbf {\bibinfo {volume} {47}},\ \bibinfo
  {pages} {16419} (\bibinfo {year} {1993})}\BibitemShut {NoStop}%
\bibitem [{\citenamefont {Thorngren}\ and\ \citenamefont
  {Else}(2018)}]{Thorngren_1612}%
  \BibitemOpen
  \bibfield  {author} {\bibinfo {author} {\bibfnamefont {R.}~\bibnamefont
  {Thorngren}}\ and\ \bibinfo {author} {\bibfnamefont {D.~V.}\ \bibnamefont
  {Else}},\ }\href {\doibase 10.1103/PhysRevX.8.011040} {\bibfield  {journal}
  {\bibinfo  {journal} {Phys. Rev. X}\ }\textbf {\bibinfo {volume} {8}},\
  \bibinfo {pages} {011040} (\bibinfo {year} {2018})},\ \Eprint
  {http://arxiv.org/abs/1612.00846} {arXiv:1612.00846} \BibitemShut {NoStop}%
\bibitem [{\citenamefont {Thorngren}()}]{Thorngren_1710}%
  \BibitemOpen
  \bibfield  {author} {\bibinfo {author} {\bibfnamefont {R.}~\bibnamefont
  {Thorngren}},\ }\href@noop {} {\ }\Eprint {http://arxiv.org/abs/1710.02545}
  {arXiv:1710.02545} \BibitemShut {NoStop}%
\bibitem [{\citenamefont {Cordova}\ \emph
  {et~al.}(2020{\natexlab{a}})\citenamefont {Cordova}, \citenamefont {Freed},
  \citenamefont {Lam},\ and\ \citenamefont {Seiberg}}]{Cordova_1905_a}%
  \BibitemOpen
  \bibfield  {author} {\bibinfo {author} {\bibfnamefont {C.}~\bibnamefont
  {Cordova}}, \bibinfo {author} {\bibfnamefont {D.}~\bibnamefont {Freed}},
  \bibinfo {author} {\bibfnamefont {H.~T.}\ \bibnamefont {Lam}}, \ and\
  \bibinfo {author} {\bibfnamefont {N.}~\bibnamefont {Seiberg}},\ }\href
  {\doibase 10.21468/SciPostPhys.8.1.001} {\bibfield  {journal} {\bibinfo
  {journal} {SciPost Physics}\ }\textbf {\bibinfo {volume} {8}},\ \bibinfo
  {pages} {001} (\bibinfo {year} {2020}{\natexlab{a}})},\ \Eprint
  {http://arxiv.org/abs/1905.09315} {arXiv:1905.09315} \BibitemShut {NoStop}%
\bibitem [{\citenamefont {Cordova}\ \emph
  {et~al.}(2020{\natexlab{b}})\citenamefont {Cordova}, \citenamefont {Freed},
  \citenamefont {Lam},\ and\ \citenamefont {Seiberg}}]{Cordova_1905_b}%
  \BibitemOpen
  \bibfield  {author} {\bibinfo {author} {\bibfnamefont {C.}~\bibnamefont
  {Cordova}}, \bibinfo {author} {\bibfnamefont {D.}~\bibnamefont {Freed}},
  \bibinfo {author} {\bibfnamefont {H.~T.}\ \bibnamefont {Lam}}, \ and\
  \bibinfo {author} {\bibfnamefont {N.}~\bibnamefont {Seiberg}},\ }\href
  {\doibase 10.21468/SciPostPhys.8.1.002} {\bibfield  {journal} {\bibinfo
  {journal} {SciPost Physics}\ }\textbf {\bibinfo {volume} {8}},\ \bibinfo
  {pages} {002} (\bibinfo {year} {2020}{\natexlab{b}})},\ \Eprint
  {http://arxiv.org/abs/1905.13361} {arXiv:1905.13361} \BibitemShut {NoStop}%
\bibitem [{\citenamefont {Kapustin}\ and\ \citenamefont
  {Spodyneiko}(2020)}]{Kapustin_2001}%
  \BibitemOpen
  \bibfield  {author} {\bibinfo {author} {\bibfnamefont {A.}~\bibnamefont
  {Kapustin}}\ and\ \bibinfo {author} {\bibfnamefont {L.}~\bibnamefont
  {Spodyneiko}},\ }\href {\doibase 10.1103/PhysRevB.101.235130} {\bibfield
  {journal} {\bibinfo  {journal} {Phys. Rev. B}\ }\textbf {\bibinfo {volume}
  {101}},\ \bibinfo {pages} {235130} (\bibinfo {year} {2020})},\ \Eprint
  {http://arxiv.org/abs/2001.03454} {arXiv:2001.03454} \BibitemShut {NoStop}%
\bibitem [{\citenamefont {Kapustin}\ and\ \citenamefont
  {Spodyneiko}()}]{Kapustin_2003}%
  \BibitemOpen
  \bibfield  {author} {\bibinfo {author} {\bibfnamefont {A.}~\bibnamefont
  {Kapustin}}\ and\ \bibinfo {author} {\bibfnamefont {L.}~\bibnamefont
  {Spodyneiko}},\ }\href@noop {} {\ }\Eprint {http://arxiv.org/abs/2003.09519}
  {arXiv:2003.09519} \BibitemShut {NoStop}%
\bibitem [{\citenamefont {Hsin}\ \emph {et~al.}(2020)\citenamefont {Hsin},
  \citenamefont {Kapustin},\ and\ \citenamefont {Thorngren}}]{Hsin_2004}%
  \BibitemOpen
  \bibfield  {author} {\bibinfo {author} {\bibfnamefont {P.-S.}\ \bibnamefont
  {Hsin}}, \bibinfo {author} {\bibfnamefont {A.}~\bibnamefont {Kapustin}}, \
  and\ \bibinfo {author} {\bibfnamefont {R.}~\bibnamefont {Thorngren}},\ }\href
  {\doibase 10.1103/PhysRevB.102.245113} {\bibfield  {journal} {\bibinfo
  {journal} {Phys. Rev. B}\ }\textbf {\bibinfo {volume} {102}},\ \bibinfo
  {pages} {245113} (\bibinfo {year} {2020})},\ \Eprint
  {http://arxiv.org/abs/2004.10758} {arXiv:2004.10758} \BibitemShut {NoStop}%
\bibitem [{\citenamefont {Hatcher}(2001)}]{Hatcher}%
  \BibitemOpen
  \bibfield  {author} {\bibinfo {author} {\bibfnamefont {A.}~\bibnamefont
  {Hatcher}},\ }\href@noop {} {\emph {\bibinfo {title} {Algebraic topology}}}\
  (\bibinfo  {publisher} {Cambridge University Press},\ \bibinfo {address}
  {Cambridge},\ \bibinfo {year} {2001})\BibitemShut {NoStop}%
\bibitem [{\citenamefont {Mermin}(1979)}]{Mermin_1979}%
  \BibitemOpen
  \bibfield  {author} {\bibinfo {author} {\bibfnamefont {N.~D.}\ \bibnamefont
  {Mermin}},\ }\href {\doibase 10.1103/RevModPhys.51.591} {\bibfield  {journal}
  {\bibinfo  {journal} {Rev. Mod. Phys.}\ }\textbf {\bibinfo {volume} {51}},\
  \bibinfo {pages} {591} (\bibinfo {year} {1979})}\BibitemShut {NoStop}%
\bibitem [{\citenamefont {Atiyah}\ and\ \citenamefont
  {Jones}(1978)}]{Atiyah_1978}%
  \BibitemOpen
  \bibfield  {author} {\bibinfo {author} {\bibfnamefont {M.~F.}\ \bibnamefont
  {Atiyah}}\ and\ \bibinfo {author} {\bibfnamefont {J.~D.~S.}\ \bibnamefont
  {Jones}},\ }\href {\doibase 10.1007/bf01609489} {\bibfield  {journal}
  {\bibinfo  {journal} {Commun. Math. Phys.}\ }\textbf {\bibinfo {volume}
  {61}},\ \bibinfo {pages} {97} (\bibinfo {year} {1978})}\BibitemShut {NoStop}%
\bibitem [{\citenamefont {Wilczek}\ and\ \citenamefont
  {Zee}(1983)}]{Wilczek_1983}%
  \BibitemOpen
  \bibfield  {author} {\bibinfo {author} {\bibfnamefont {F.}~\bibnamefont
  {Wilczek}}\ and\ \bibinfo {author} {\bibfnamefont {A.}~\bibnamefont {Zee}},\
  }\href {\doibase 10.1103/PhysRevLett.51.2250} {\bibfield  {journal} {\bibinfo
   {journal} {Phys. Rev. Lett.}\ }\textbf {\bibinfo {volume} {51}},\ \bibinfo
  {pages} {2250} (\bibinfo {year} {1983})}\BibitemShut {NoStop}%
\bibitem [{\citenamefont {Volovik}\ and\ \citenamefont
  {Yakovenko}(1989)}]{Volovik_1989}%
  \BibitemOpen
  \bibfield  {author} {\bibinfo {author} {\bibfnamefont {G.~E.}\ \bibnamefont
  {Volovik}}\ and\ \bibinfo {author} {\bibfnamefont {V.~M.}\ \bibnamefont
  {Yakovenko}},\ }\href {\doibase 10.1088/0953-8984/1/31/025} {\bibfield
  {journal} {\bibinfo  {journal} {J. Phys.: Condens. Matter}\ }\textbf
  {\bibinfo {volume} {1}},\ \bibinfo {pages} {5263} (\bibinfo {year}
  {1989})}\BibitemShut {NoStop}%
\bibitem [{\citenamefont {Wang}\ and\ \citenamefont
  {Senthil}(2014)}]{Wang_1401}%
  \BibitemOpen
  \bibfield  {author} {\bibinfo {author} {\bibfnamefont {C.}~\bibnamefont
  {Wang}}\ and\ \bibinfo {author} {\bibfnamefont {T.}~\bibnamefont {Senthil}},\
  }\href {\doibase 10.1103/PhysRevB.89.195124} {\bibfield  {journal} {\bibinfo
  {journal} {Phys. Rev. B}\ }\textbf {\bibinfo {volume} {89}},\ \bibinfo
  {pages} {195124} (\bibinfo {year} {2014})},\ \Eprint
  {http://arxiv.org/abs/1401.1142} {arXiv:1401.1142} \BibitemShut {NoStop}%
\bibitem [{\citenamefont {Witten}(1984)}]{Witten_1984}%
  \BibitemOpen
  \bibfield  {author} {\bibinfo {author} {\bibfnamefont {E.}~\bibnamefont
  {Witten}},\ }\href {\doibase 10.1007/BF01215276} {\bibfield  {journal}
  {\bibinfo  {journal} {Commun. Math. Phys.}\ }\textbf {\bibinfo {volume}
  {92}},\ \bibinfo {pages} {455} (\bibinfo {year} {1984})}\BibitemShut
  {NoStop}%
\bibitem [{\citenamefont {Liu}\ and\ \citenamefont {Wen}(2013)}]{Liu_1205}%
  \BibitemOpen
  \bibfield  {author} {\bibinfo {author} {\bibfnamefont {Z.-X.}\ \bibnamefont
  {Liu}}\ and\ \bibinfo {author} {\bibfnamefont {X.-G.}\ \bibnamefont {Wen}},\
  }\href {\doibase 10.1103/PhysRevLett.110.067205} {\bibfield  {journal}
  {\bibinfo  {journal} {Phys. Rev. Lett.}\ }\textbf {\bibinfo {volume} {110}},\
  \bibinfo {pages} {067205} (\bibinfo {year} {2013})},\ \Eprint
  {http://arxiv.org/abs/1205.7024} {arXiv:1205.7024} \BibitemShut {NoStop}%
\bibitem [{\citenamefont {Else}\ \emph {et~al.}()\citenamefont {Else},
  \citenamefont {Huang}, \citenamefont {Prem},\ and\ \citenamefont
  {Gromov}}]{ToAppear}%
  \BibitemOpen
  \bibfield  {author} {\bibinfo {author} {\bibfnamefont {D.~V.}\ \bibnamefont
  {Else}}, \bibinfo {author} {\bibfnamefont {S.-J.}\ \bibnamefont {Huang}},
  \bibinfo {author} {\bibfnamefont {A.}~\bibnamefont {Prem}}, \ and\ \bibinfo
  {author} {\bibfnamefont {A.}~\bibnamefont {Gromov}},\ }\href@noop {} {\
  }\Eprint {http://arxiv.org/abs/2103.13393} {arXiv:2103.13393} \BibitemShut
  {NoStop}%
\bibitem [{\citenamefont {Kitaev}(2006)}]{Kitaev_0506}%
  \BibitemOpen
  \bibfield  {author} {\bibinfo {author} {\bibfnamefont {A.}~\bibnamefont
  {Kitaev}},\ }\href {\doibase 10.1016/j.aop.2005.10.005} {\bibfield  {journal}
  {\bibinfo  {journal} {Ann. Phys.}\ }\textbf {\bibinfo {volume} {321}},\
  \bibinfo {pages} {2} (\bibinfo {year} {2006})},\ \Eprint
  {http://arxiv.org/abs/cond-mat/0506438} {arXiv:cond-mat/0506438} \BibitemShut
  {NoStop}%
\bibitem [{\citenamefont {Kitaev}(2015)}]{KitaevIPAM}%
  \BibitemOpen
  \bibfield  {author} {\bibinfo {author} {\bibfnamefont {A.}~\bibnamefont
  {Kitaev}},\ }\href@noop {} {} (\bibinfo {year} {2015}),\ \bibinfo {note}
  {{I}nstitute for Pure Applied Mathematics, UCLA.
  \url{http://www.ipam.ucla.edu/abstract/?tid=12389\&pcode=STQ2015
  }}\BibitemShut {NoStop}%
\end{thebibliography}%
\end{document}